\documentstyle[preprint,aps,psfig,epsf]{revtex}

\begin{document}
\tightenlines

\title{Thermal denaturation and folding rates of single domain proteins:
size matters}
\author{Mai Suan Li$^1$, D. K. Klimov$^2$ and D. Thirumalai$^2$}

\address{$^1$Institute of Physics, Polish Academy of Sciences,
Al. Lotnikow 32/46, 02-668 Warsaw, Poland\\
$^2$Institute for Physical
Science and Technology, University of Maryland, College Park, MD 20742 }
\maketitle

\begin{abstract}
We analyze the dependence of thermal denaturation transition and
folding rates of globular proteins on the number of amino acid
residues, $N$. Using lattice Go models we show that $\Delta T/T_F \sim
N^{-1}$, where $T_F$ is the folding transition temperature and $\Delta
T$ is the transition  width computed using the temperature dependence of
the order parameter that distinguishes between the unfolded state and the
native basin of attraction. This finding is consistent with finite
size effects expected for the  systems undergoing a phase transition
from a disordered to an ordered phase. The dependence of the folding
rates on  $N$ for lattice models and the  dataset of 57 proteins and
peptides shows that $k_F \simeq  k_F^0 exp(-CN^\beta)$ with $0<
\beta \le 2/3$  provides a good fit, where $C$ is a $\beta$-dependent
constant.  
We find that $k_F \simeq  k_F^0 exp(-1.1N^{\frac{1}{2}})$ with
an average (over the dataset of proteins) 
$k_F^0 \approx (0.4\mu s)^{-1}$, 
can estimate optimal protein folding rates, to within an order
of magnitude in most cases. By using this fit for a set of proteins with
$\beta$-sheet topology we find that $k_F^0\approx  k_U^0$, 
the prefactor for unfolding. The maximum ratio of $k_U^0/k_F^0 \approx 10$
for this class of proteins.
\end{abstract}

\section{Introduction} 

Deciphering the factors that determine the foldability of protein sequences 
\cite{Onuchic97,Thirumalai99,Dill97,Mirny01,Boneau01}
is an important problem from the perspective of protein design,
protein structure prediction, and 
{\em in vitro} and {\em in vivo} protein folding. Foldability refers 
both to the folding rate, $k_F$, and thermodynamics of the transition from
the ensemble of unfolded states ({\bf U}) to the native state or, more
precisely, to 
the native basin of attraction ({\bf NBA}). 
Folding rates and the associated equilibrium characteristics depend
on intrinsic factors (sequence and topology) as well as  on 
external conditions (pH, temperature, salt concentration, 
and viscosity). Variation in external conditions
can not only alter the rates,
but also the mechanism of folding. Despite this obvious fact
most of the studies have been focused  on the dependence of
$k_F$
solely on the characteristics of the native states as described by the
crystal (or NMR) structures.

The role of finite size effects on the thermodynamics of protein
folding has received very little attention.
The emphasis  on the cooperativity of
the transition from {\bf U} state to 
{\bf NBA} seems to have precluded consideration of the role
of $N$, the number of residues in a sequence. 
This transition, for apparent two-state folders, has all the
hallmarks of (weak) first-order phase transition. The highly 
cooperative  {\bf U$\leftrightarrow$NBA} transition
has lead some authors to suggest that 
there is no evidence that partially structured states 
contribute to the thermodynamic properties of
proteins. Computational studies have shown \cite{Klimov00a} that 
in $\beta$-hairpin
forming sequence from the C-terminus of GB1 protein structure is
acquired over a finite range of temperatures, even though the overall
folding can be described as a broad "two-state"
transition \cite{Munoz97}. Experiments on refolding of barnase have also 
suggested that
structure is lost incrementally upon temperature induced
unfolding \cite{Laksh01NSB}. 
Direct temperature dependence of structure formation
in leucine zipper using  one dimensional NMR
experiments has established that melting
temperature varies across the structure \cite{Holtzer97Biophys}. 
Although the variations occur
over a relatively narrow range of temperatures, it is clear from these
experiments that because of the finite size of proteins
partially folded structures contribute to folding thermodynamics. 
These observations warrant an examination of finite size effect on the
{\bf U $\leftrightarrow$ NBA} transition. 
Building on  our previous study \cite{Klimov02},
 we further investigate the role of $N$ in thermal denaturation 
using lattice models of proteins.

It has been noted \cite{Plaxco98} that 
$k_F$  correlates well
with the relative contact order (RCO), which measures the proximity of
side chain contacts in the folded state. The notion that protein folding
is initiated with
residues forming local structures and, thus, is  determined by their
proximity along the sequence constitutes the
basis of the hierarchical folding mechanism \cite{Baldwin99}. 
Thus, in retrospect, the correlation between the RCO and $k_F$
is not entirely unanticipated, especially in $\alpha$-helical
proteins. Although RCO is an important indicator of the folding rates, it
should be pointed out that there is little correlation between RCO and
$k_F$ for 
proteins with $\beta$-sheet topology. Clarke {\em et al} 
\cite{Clarke99} showed that
neither $k_F$ nor the unfolding rates $k_U$ 
correlate with RCO for a class of
$\beta$-sheet proteins belonging to the immunoglobulin (Ig) fold. The RCO
for the 6 proteins examined \cite{Clarke99} is in the very narrow range
($0.17 \leq \text{RCO} \leq 0.20$). Nevertheless, 
the refolding rates for these proteins
vary  by a factor of 800. More recently, Clarke and coworkers have shown that
for a number of Ig domains from the muscle protein titin 
$k_F$ can vary by over four orders of magnitude \cite{Clarke02JMB},
although 
their RCO values are expected to be nearly the same. These studies show that
factors besides RCO play an important role in the determination of $k_F$.

Surprisingly, it was initially suggested \cite{Plaxco98} that neither
stability nor the size
of proteins plays a role in determining $k_F$. These
counterintuitive observations contradict several theoretical
\cite{Thirumalai95,Wolynes97,Finkelstein97,Gutin96} and a few
experimental
studies \cite{Clarke99}. More careful examination of the database of well
characterized proteins has shown that, although there
are  exceptions  \cite{Perl98},
stability is an important factor
that determines $k_F$ \cite{Clarke99,Dinner01}. Recently, several studies
\cite{Galzitskaya03,Ivankov03} have concluded
that $N$, the number of amino acids, must also 
play an important role in determining $k_F$.
In this paper we examine the dependence of rates as well as
thermal denaturation of single domain proteins on $N$. 

Beginning with the paper by one of us  \cite{Thirumalai95} a number of
theoretical studies \cite{Wolynes97,Finkelstein97,Gutin96}
have predicted that $N$ should play a significant
role in controlling  $k_F$. Given that polypeptide chains are
heteropolymers we expect that their relaxation times in both the
folded and unfolded states  must depend on $N$.
Theoretical studies \cite{Thirumalai95,Wolynes97}
suggest that the dependence
of  $k_F$ on $N$ is dictated by the
interplay of three characteristic temperatures of the polypeptide chain,
namely, $T_F$ (the folding transition temperature), $T_\theta$
(the collapse transition temperature), and
$T_g$ (the glass transition temperature). It
appears that in most experiments the external conditions are such that fastest
folding is observed near the "tricritical" point, where
$T_F \approx T_{\theta}$ in accord with the prediction by Camacho and
Thirumalai \cite{Camacho93}. For near optimal folding, as it
may be the case for 
minimally frustrated sequences, it has been argued
that
\begin{equation}
\text{ln}(k_F/k_F^0) \; \;  \sim \; \; \alpha \text{ln} N,
\end{equation}
where $\alpha\approx 4$ \cite{Thirumalai95}
and $k_F^0$ is an undetermined prefactor. 
For artificial Go models $ \alpha\approx
3$ \cite{Gutin96,Cieplak99}. On the other hand, due
to topological frustration, even the sequences following two-state
kinetics have a rough energy landscape. In this case
\begin{equation}
\text{ln}(k_F/k_F^0) \; \;  \sim \; \; N^\beta. 
\end{equation}
The value of $\beta$ has been suggested to be less than unity and is
probably in the range $0.5\le \beta \le \frac23$ 
\cite{Thirumalai95,Wolynes97,Finkelstein97}. Given the limited
range of $N$ for single domain proteins 
it is difficult (see below) to determine $\beta$ precisely.

To probe finite size effects on thermally induced folding
we have performed Monte Carlo simulations
using Go lattice models. These results are used to quantitatively
establish the effect of finite $N$ on rounding the 
{\bf U $\leftrightarrow$ NBA} transition. A dataset of
proteins, for which   $k_F$ is available, is used to draw lessons on the
dependence of $k_F$ on $N$.
Using these results we show that unambiguous determination of 
$\beta$ is not possible. However, 
we argue that the $N$ dependence given in Eq. (2)
is useful in analyzing the experimental data. 
As a byproduct of this work we also provide estimates of the folding
and unfolding prefactors, $k_F^0$ and  $k_U^0$.


\section{Models and methods}

For the numerical simulations we represent a polypeptide chain using
lattice Go model without side chains. The energy of a conformation 
\begin{equation}
E \, = \,  \sum_{i<j} \epsilon_{ij} \delta_{r_{ij},a},
\label{H_noSC}
\end{equation}
where $a$ is a lattice spacing, $r_{ij}$ is the distance between
non-bonded beads $i$ and $j$,  and the contact energies $\epsilon_{ij}$
are chosen 
to be -1 for native contacts and 0 for non-native ones. 
Go models are useful in exploring general physical principles that 
govern protein folding under the condition of
marginal stability of the native state \cite{Chan1,Chan2}. The sequences
were selected by a standard sequence space Monte Carlo algorithm, which
maximizes the Z-score for a given target structure. The target
structures for each $N$ were chosen to be maximally compact. For
example, for $N=18$ and $N=80$ the native structures  
occupy the vertexes of 3x3x2 and 4x4x5
cubes, respectively. 

The thermodynamics of folding were determined using
Monte Carlo simulations based on MS3 move
set \cite{Betancourt,Betancourt02,Li02,Li02a}, 
which involves single, double and triple
bead moves. Because this move set involves multiparticle updates, it is
much more efficient compared to the standard
move set \cite{Li02,Li02a,Hilhorst75}. 
The thermodynamic properties of the sequences are calculated using the multiple
histogram method \cite{Ferrenberg89}. Typical number of Monte Carlo
trajectories used to collect histograms is 50-100 depending on $N$.
The free energy is calculated as
a function of the 
number of native contacts $Q$,  which is  treated  as
an approximate reaction
coordinate for Go models. This allows us to estimate the dependence of
folding and unfolding free energy barriers on $N$.

For lattice models 
the structural similarity with the native conformation is measured by the
overlap function \cite{Camacho93}
\begin{eqnarray}
\chi = 1 - \frac{1}{N^2-3N+2} \sum_{i<j+1}^N \delta (r_{ij}-r_{ij}^0),
\label{chi_eq}
\end{eqnarray}
where the superscript $0$ refers to the native state.
The folding temperature $T_F$ is defined as a temperature at which
$d<\chi >/dT$ is maximum and the transition width $\Delta T$ is 
defined as the full
width at half maximum of $d<\chi >/dT$ at $T = T_F$.


\section{Results }

\begin{figure}
\epsfxsize=7in
\vspace{2in}
\centerline{\epsffile{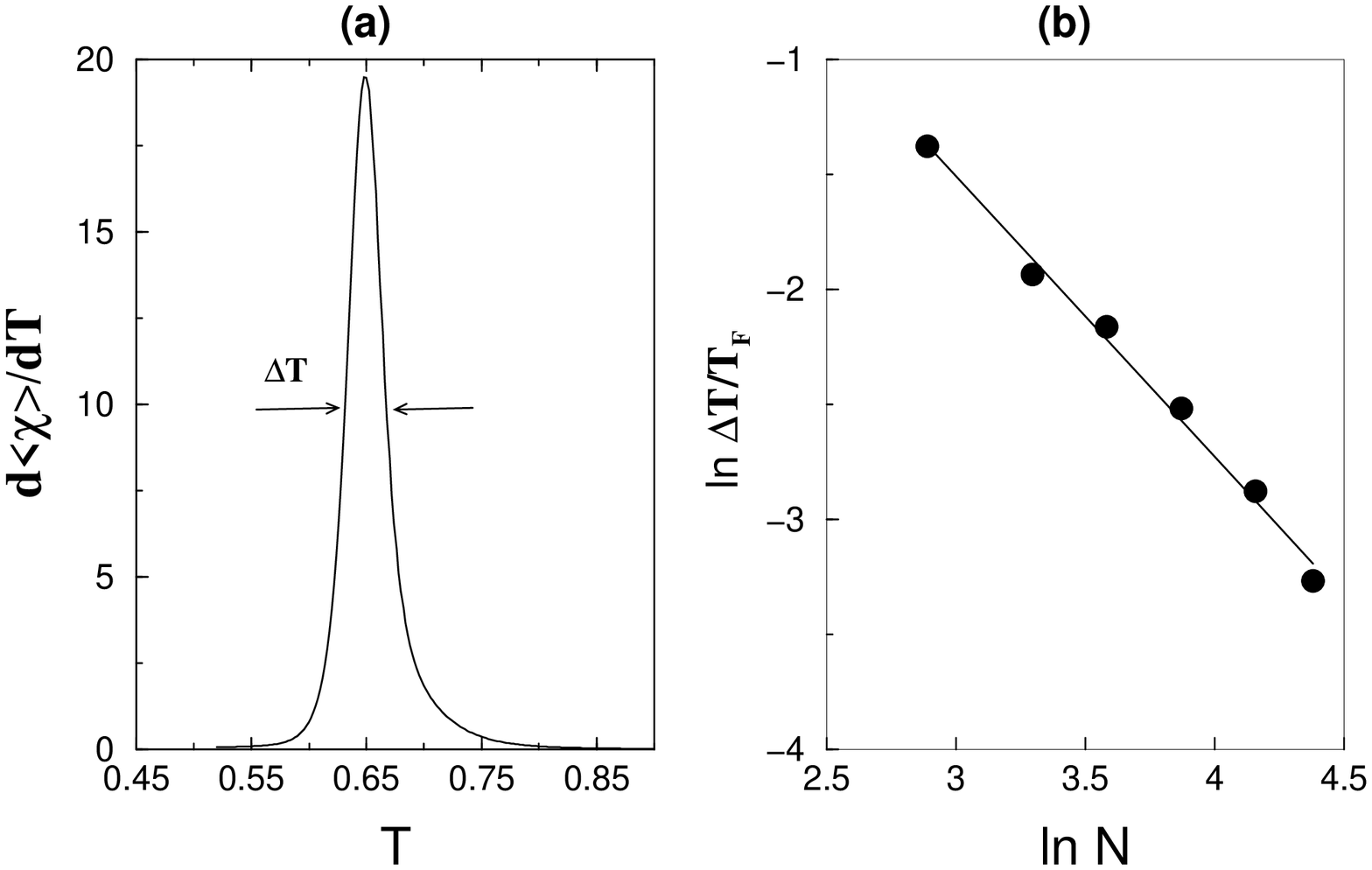}}
\vspace{0.1in}
\caption{}
\label{chi_fig}
\end{figure}

\noindent
{\bf A. Finite size effects in 
thermal denaturation:}  The transition width
$\Delta T$ is obtained from the temperature
dependence of $d<\chi>/dT$ (see Fig. \ref{chi_fig}a for an example).
For all the
sequences considered here $T_F\approx T_\theta$.
For finite size systems 
the {\bf U $\leftrightarrow$ NBA} transition is expected to be
rounded. The rounded nature of the transition which has been seen in
simulations,
is reflected in the temperature dependence of 
$d<\chi >/dT$ (Fig. \ref{chi_fig}a). More importantly, we expect
$\Delta T/T_F$ to  scale as
\begin{equation}
\frac{\Delta T}{T_F} \sim N^{-1}.
\label{deltaT}
\end{equation}
The data for lattice Go models show that
$\Delta T/T_F \sim N^{-\lambda}$ with $\lambda = 1.2\pm 0.1$
(Fig. \ref{chi_fig}b).
The small deviation from the expected theoretical result
(Eq. (\ref{deltaT})) may be a consequence of the relatively small 
$N \le 80$ in the sample. For small values of $N$ the native state
does not have a well-defined core. As a result fluctuations are
relatively large, which may explain the observed deviation.  Analysis
of the experimental data indeed shows that (Eq. (\ref{deltaT})) is
obeyed with great precision \cite{Klimov02,Li03}.

\noindent

{\bf B: $N$ dependence of folding and unfolding barrier heights at $T_F$
for Go models}. To compute the free energy folding barriers,
$\Delta F^\ddagger_F (\simeq \Delta F^\ddagger_U$, the unfolding
barrier, at $T_F$) it is necessary to define a 
reaction coordinate. The precise reaction coordinate for
a multi-dimensional process such as protein folding is difficult to
ascertain. However, Onuchic and coworkers \cite{Nymeyer98PNAS} 
have argued that, for minimally 
frustrated systems such as the Go models, the fraction
of native contact $Q$ may be appropriate. Accordingly, we have computed
$F(Q)$ for about 80 sequences with $N$ ranging from 
18 to 80. This is the largest number of sequences used so far to test the
expected scaling of $\Delta F^\ddagger_F$ and $\Delta
F^\ddagger_U$. At $T_F$, 
$\tau^0_F \text{exp}(\Delta F^\ddagger_F/k_BT_F) = 
\tau^0_U \text{exp}(\Delta F^\ddagger_U/k_BT_F)$. Because it is not 
obvious that $\tau ^0_F \approx \tau ^0_U$, 
$\Delta F^\ddagger_F$ and $\Delta F^\ddagger_U$ may, 
in principle,  exhibit different scaling behavior with $N$. 

\begin{figure}
\epsfxsize=8in
\vspace{3in}
\centerline{\epsffile{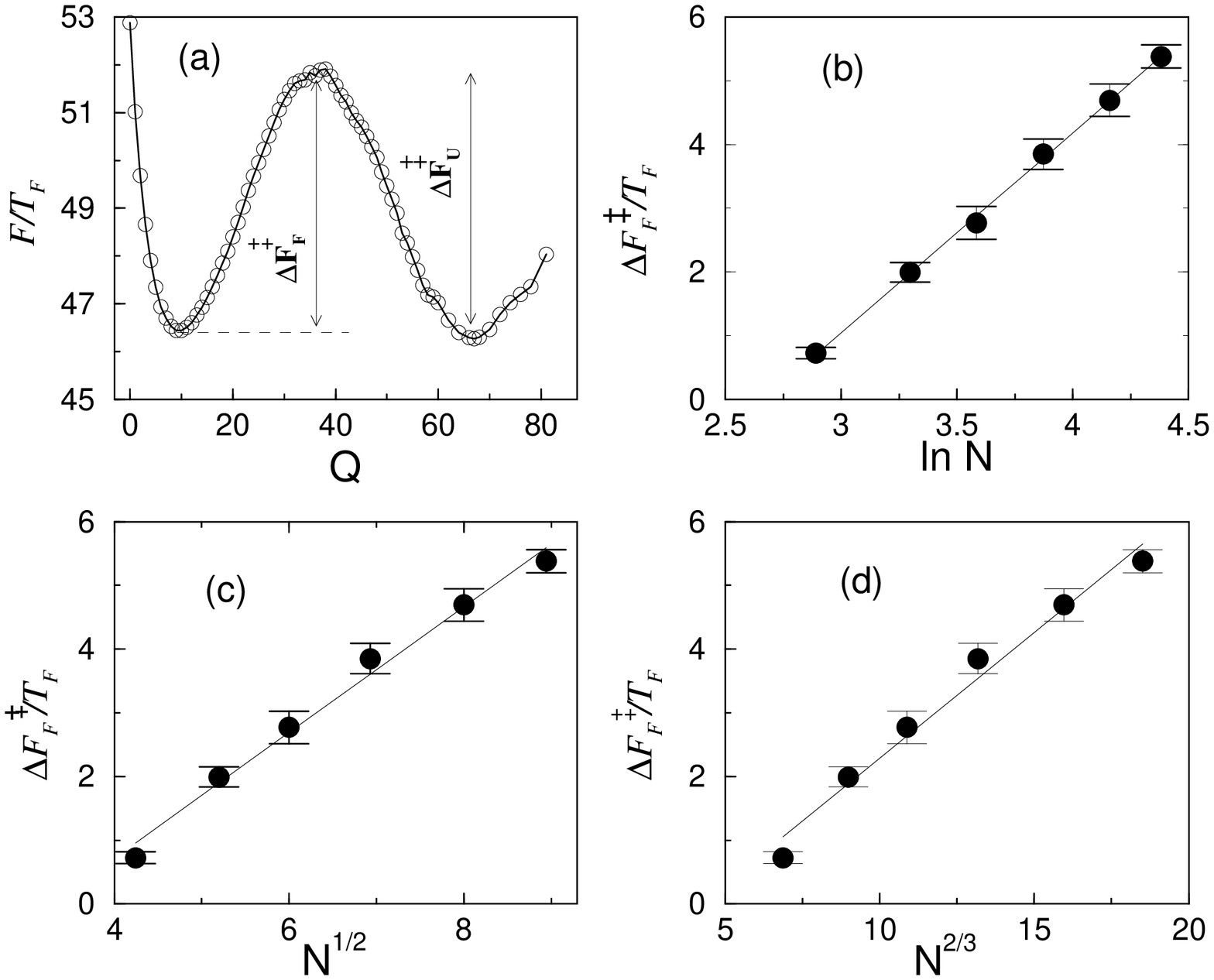}}
\vspace{0.2in}
\caption{}
\label{FreeE_scaling}
\end{figure}

From the typical free energy profile $F(Q)$  
(Fig \ref{FreeE_scaling}a) we computed $\Delta F^\ddagger_F$
and $\Delta F^\ddagger _U$. The variation
of $\Delta F^\ddagger_F/k_BT$ as a function of ln$N$, $N^{1/2}$ and
$N^{2/3}$ for 
the Go sequences plotted in Fig. \ref{FreeE_scaling}b,c,d, respectively,
shows that all three fits quantitatively reproduce the simulation
results. However, we argue below using the analysis of experimental data that
$\Delta F^\ddagger_F \sim \text{ln}N$ is not viable. 
Based on experimental
estimates of $\tau_F^0$ and $\tau^0_U$ we find that 
$\Delta F^\ddagger_F \sim N^{1/2}$ provides the best physically
acceptable representation of the data. From the lattice model computations
we find  
$\Delta F^\ddagger_F$ and $\Delta F^\ddagger_U$ have the same 
dependence on $N$, which
implies that $\tau_F^0 \approx \tau_U^0$.

\noindent
{\bf C. Chain length dependence of folding rates: } The RCO, which
is a characteristic of the native topology of proteins,
is \cite{Plaxco98}
\begin{equation}
RCO = \frac{\sum_{i,j}\Delta_{ij} |i-j| }{N \sum_{i,j}\Delta_{ij}},
\label{RCO}
\end{equation}
where $|i-j|$ is the sequence separation between the residues $i$ and
$j$ and $\Delta_{ij}$ is unity, if $i$ and $j$ form a native contact, or
zero, otherwise. The observed correlation between RCO and
ln $k_F$ suggests that folding is most rapid, if the native state 
has a large fraction of local contacts. The importance of RCO
is based on the sound physical idea that residues local in
sequence space tend to form interactions early in the folding process
and, if these substructures
can "coherently" add to produce the folded structure, efficient folding may be
realized. However, almost all proteins are stabilized by a
sizable fraction of long-ranged (non-local)  contacts. This suggests
that ln $k_F$ may also depend on other factors (for example, 
stability \cite{Dinner01} and $N$)
besides RCO. Lattice model simulations \cite{Klimov98a} and 
experiments \cite{Clarke99} have shown that native state stability is also a
contributing factor to refolding rates.

Depending on the extent of energy frustration 
one of us  suggested \cite{Thirumalai95} that $k_F \sim N^\alpha$ (for
optimized sequences) or $k_F \sim \text{exp}(-C_1 N^\frac12)$, where
$C_1$ is a constant (Eqs. (1,2)).
By balancing the "bulk" free energy gain due to the formation of
a stable hydrophobic core and the surface tension cost due to interface
formation
it has been proposed \cite{Wolynes97,Finkelstein97} that for optimal folding
$k_F \sim$ exp$(-C_2 N^\frac23)$,  where $C_2$ is a constant. 
Although,  the limited range of $N$ values accessible in
proteins makes it difficult to unambiguously determine
the precise way $k_F$ decreases upon increasing $N$, it is generally agreed
that free energy barriers in proteins shall be relatively small.
Moreover, the
transition region could be broad with roughness superimposed on it. As
a result $\Delta F^\ddagger_F/k_BT_F$ is expected to grow only as
$N^\beta$ with $\beta < 1$. The sublinear growth of 
$\Delta F^{\ddagger}_F/k_BT$
with respect to $N$ naturally explains both the rapid folding
(kinetics) and marginal stability (thermodynamics) of folded states 
of proteins.

Recently, Koga and Takada \cite{Koga01}
have computed folding rates for 18 proteins
using $C_\alpha$-Go models. They fit the data using $k_F \sim 
\text{exp}(-C_3 RCO\times N^\beta)$ with $\beta=0.607\pm0.179$
and $C_3$ is a constant. Within the error bar of
their fit it is impossible to distinguish between $\beta =0.5$ or
2/3. Their results showed, as argued on theoretical grounds, that
$\beta < 1$. In addition, due to the possibility that
RCO decreases with $N$ \cite{Ivankov03} it is likely that the actual value of
$\beta$ in \onlinecite{Koga01} is considerably smaller.
By focusing on the proteins that fold by three-state
kinetics Galzitskaya {\em et al.}
\cite{Galzitskaya03} have argued that chain length $N$ is
the major determinant of folding rates. However, they were unable to
determine the precise dependence of $k_F$ on $N$.

Ivankov {\em et el.} \cite{Ivankov03}
have reconsidered chain length
dependence of $k_F$ by analyzing experimental data for 57 proteins
(both two and three state folders) and peptides. They
suggested that ln $k_F \sim -0.44 RCO\times N + 11.15$ for the set of 57
proteins with the correlation coefficient $p= 0.74$. For this dataset of
proteins it is argued that RCO $\sim N^{-0.3}$, so that $k_F \sim
\text{exp}(-C_4 \times N^{0.7})$, where $C_4$ is a constant. 
Because there are errors in fitting RCO to a power law
decay with $N$, the indirect inference that $\beta \approx 0.7$ is not
transparent.  
To circumvent this problem we have directly examined the dependence of 
ln$k_F$ on $N$. The fit of ln$k_F$ using
the theoretically proposed models are shown in Fig. \ref{real_pro_fig}a, b
and c. The correlation coefficient for the fits
ln$k_F \sim N^{\beta}$ is nearly constant for $0 \le \beta \le 2/3$ 
and begins to 
decrease modestly for $\beta > 2/3$ (Fig. \ref{real_pro_fig}d).
We have also established 
that the folding rates in lattice Go models
can be adequately fit with $\beta$=0, 0.5, or 2/3 \cite{Li02a}. 
From this perspective alone  it is difficult to distinguish between the three
theoretical values of $\beta$ (0, 0.5, and $2/3$). 
However, we rule out ln$k_F \sim$ ln$ N$ (Fig. \ref{real_pro_fig}b)
based on the following arguments: (1) The power law fit yields 
ln$k_F = -5.5$ ln$N + 28.5$ which implies $k_F^0 \approx e^{28.5} s^{-1}
= (0.4 ps)^{-1}$. This value for the prefactor $k_F^0$ is nearly the same as
$k_BT/h \approx (0.2 ps)^{-1}$, which is reasonable for small molecules,
but is not appropriate to describe folding reactions.
(2) The value of the exponent $\alpha = 5.5$ is too large to be justified 
theoretically. Such a large value of $\alpha$ is usually indicative of an 
underlying activated process with a relatively small barrier \cite{Li02a}.

The fits to the data in Fig. (3) cannot distinguish  
the scalings of $ln k_F$ with $N^{1/2}$ or $N^{2/3}$. This is consistent with
our results presented in Fig. \ref{FreeE_scaling}.
In an attempt 
to further discriminate between $\beta = 0.5$ and $\beta = 2/3$ we 
focus on the numerical values of the prefactor $k_F^0$. 
The inverse of the prefactor $1/k_F^0$ for the $N^{1/2}$ scaling of
the barrier height is $0.4 \mu s$, whereas 
$1/k_F^0 \approx 8 \mu s$ for $N^{2/3}$ scaling 
(see caption to Fig. \ref{real_pro_fig}).
By applying Kramer's theory to describe the
{\bf U $\leftrightarrow$ 
NBA} transition  it has been argued that $\tau_0 = 1/k_F^0$ should be 
considerably greater than $h/k_BT$ 
\cite{KlimThirum97PRL,Socci96JCP,Portman01}. The range 
$0.4 \mu s \leq \tau _0 \leq 8 \mu s$ obtained from the two fits is consistent
with this expectation. Therefore, it follows that, unless a direct experimental
measurement of $k_F^0$ is made, it would be difficult to determine the
precise value of $\beta$. The goodness of fits with $\beta = 1/2$ or
$\beta = 2/3$ shows clearly that barriers to folding scale sublinearly
with $N$.

\noindent

{\bf D. Prefactors for folding and unfolding}.
There is considerable interest in obtaining a fairly accurate estimate of 
$\tau _F^0 (\sim (k_F^0)^{-1})$  at near  neutral pH and $T=25^o$C so
that the measurements of average barrier heights can be made directly.
Estimates of $\tau _F^0$ have been made using few physically
motivated arguments:

(1) Assuming that the most elementary step
in the folding process is the formation of a single tertiary contact (a loop
between two residues separated by $l$ intervening
residues) it was argued that the
speed limit for folding is about $1 \mu s$ \cite{Hagen96}. 
Because most probable loops are
predicted to form in about $1 \mu s$ \cite{Guo95}, it follows that 
$\tau _F^0 \approx 1 \mu s$.
Eaton and coworkers \cite{Hagen96,Munoz99} 
have provided additional arguments that proteins are
unlikely to fold faster than $\tau_F^0 \approx 1\mu s$.

(2) Yang and Gruebele \cite{Yang03} 
argue using refolding 
data of mutants of a helical protein $\lambda_{6-85}$ that 
$\tau _F^0 \approx 2 \mu s$. We believe that, in general,  
for the majority of proteins $\tau _F^0 \approx 2 \mu s$
should be near the upper limit
for the following reasons. Based on theories of
collapse dynamics we expect that the 80 
residue protein $\lambda_{6-85}$ becomes compact
in about $\tau _c \approx (\eta a/\gamma)N^{\theta}
\approx 1.5 \mu s$, where $\eta$ is the solvent viscosity, $a$ is the Flory 
characteristic ratio, $\gamma$ is the surface tension (50 cal/mol$\AA^2$),
$\theta \approx 2.2$
and $N=80$. This estimate is close to the folding time for
$\lambda _{6-85}$, which suggests that collapse and folding are nearly 
simultaneous for this protein. Because these two processes cannot be separated
for proteins with $N=80$ that fold in about $\approx 1 \mu s$,
it appears that one can assume that
$\tau _F^0 \approx 2 \mu s$ may be an upper bound. 
We believe that $\tau _F^0 \approx 1 \mu s$
could serve as a practical estimate for the prefactor, 
because on time scales greater
than $1 \mu s$ multiple loops
can form and collapse of the entire polypeptide chain can occur, which
could obfuscate direct determination of $\tau_F^0$.  
In arriving at these estimates for $\lambda_{6-85}$ we have assumed that 
internal viscosity does not alter folding rates appreciably.
Although, a similar observation has been made for refolding of protein L 
\cite{Plaxco99} and CspB \cite{Schmid97PNAS} 
it is unclear how important internal viscosity of proteins is in the
determination of $\tau _F^0$ \cite{Portman01}. 
In addition, external conditions can  alter $\tau^0_F$. Thus, 
$\tau^0_F \approx 1 \mu s$ should be taken merely as a useful
estimate for the prefactor. 

The dependence of ln $k_F$ on $N$ (Fig. (3)) allows us to estimate
$\tau _F^0$ and $\tau _U^0$
using the experimental data for proteins that are not 
fully represented in Fig. (3). We use refolding rates for
several $\beta$-sheet proteins to estimate
$\tau _F^0$ (Table 1). Assuming  $N^{1/2}$
scaling we find that $\tau_F^0$ is in the range $(0.1-18)\mu s$. Except for
$\tau_F^0\simeq 18\mu s$ obtained for twitchin (TWIg18') with low
native state stability the average value of the prefactor is
$\overline{\tau_F^0}\approx 3.5\mu s$. If we use $\tau_F^0\simeq \tau_F
exp(-0.36N^\frac23)$ (Fig. (3)), we find $2 \mu s
\lesssim \tau_F^0 \lesssim 400\mu s$ (Table 1). 
For four immunoglobulin proteins
with the exception of FNfn10 (Table 1)  
the estimated values of $\tau_F^0$ using the
$N^\frac23$ scaling for the barrier height seem too large. 
Thus, $\tau_F^0$ appears to be in the neighborhood of
few $\mu s$ for the $\beta$-sheet proteins and for the 
$\alpha$-helical protein $ \lambda_{6-85}$.

Another question of interest is whether $\tau_F^0 \approx \tau_U^0$?
Using lattice model simulations we have previously argued that the
unfolding and folding prefactors are similar \cite{Klimov98a}. 
This conclusion was reached
using the number of native contacts $Q$ as a  reaction coordinate. It
is unclear whether this result is a consequence of our choice of 
the  reaction coordinate.  The results in Fig. (3) and the measured
unfolding rates in Table 1 allow us to directly estimate
\begin{equation}
\tau_U^0 \simeq \tau_U \exp{[-(1.1 N^{1/2}+\beta \Delta G)]},
\end{equation}
where $\tau_U$ is the unfolding time, $\Delta G$ is the free energy of
stability of the native state, and $\beta=(k_BT)^{-1}$.  With the
exception of TWIg18' the ratio $\tau_U^0/\tau_F^0 < 1$ and is in
the range $0.1 \lesssim \tau_U^0/\tau_F^0 \lesssim 1.0$. For this
class of proteins the maximum value of $\tau_F^0/\tau_U^0 \lesssim 10$
(Table 1). Similar conclusions have been drawn for $\alpha$-helical
proteins as well. Thus, it appears that   $\tau_U^0 \approx \tau_F^0$.

\section{Conclusions}

\begin{figure}
\epsfxsize=7in
\vspace{2in}
\centerline{\epsffile{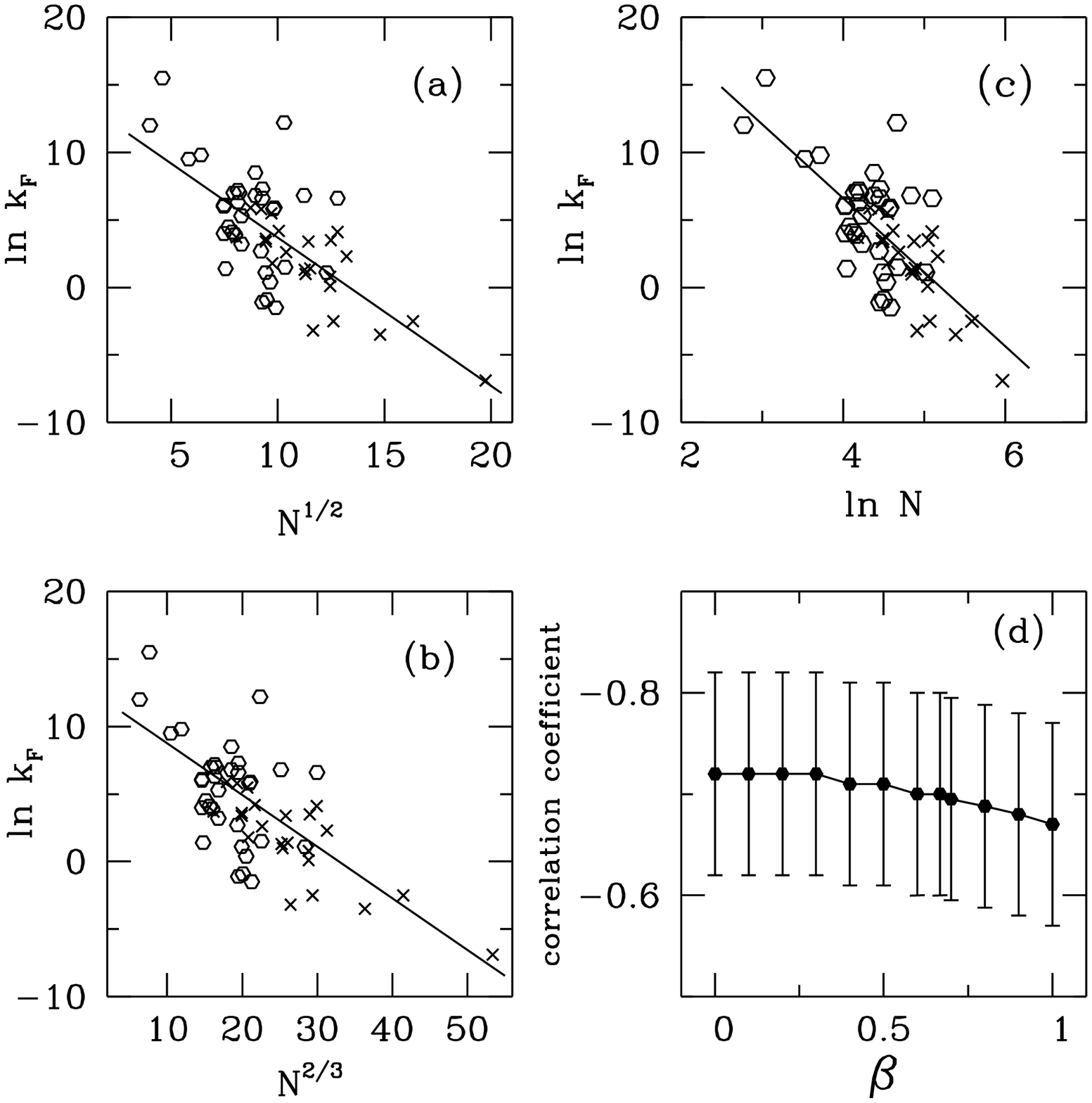}}
\vspace{0.2in}
\caption{}
\label{real_pro_fig}
\end{figure}

In this article we have considered finite size effects in thermal
denaturation and folding kinetics. We have established using lattice
models  that the rounded transition as quantified by $\Delta T/T_F$
obeys the expected scaling (Eq. (\ref{deltaT})). This is in accord with the
earlier analysis of the experimental data
\cite{Klimov02}, which further suggests that
qualitative features of folding transition can be gleaned using
lattice models. Unlike the case of thermal denaturation
the situation is far more ambiguous when the
scaling  of $k_F$ with $N$ is considered. The dependence of ln $k_F$
on $N$ does not match the quality of 
correlation noted for thermodynamics. If we
delete the fastest folding proteins and peptides and the slowest
folding proteins from the dataset in Fig. \ref{real_pro_fig}, 
the correlation coefficient becomes
considerably worse  ($\approx 0.56$) regardless of the scaling
($\beta =0.5$ or 2/3) used. Nevertheless, the inclusion of the $N$
dependence does improve the correlation between $ln k_F$
on RCO \cite{Ivankov03}. Using the expected values (from a number of
unrelated studies) for the prefactor, we
suggest that the $N^{1/2}$ scaling  for barrier height $\Delta
F^\ddagger/k_BT$ may be useful in making order of magnitude estimates
of refolding rates. This scaling also implies that 
the energy landscape of two-state proteins is rugged. The energy
scale for roughness may be of order of a few $k_B T$.

We are grateful to A. Finkelstein for permission to use the data in
Ref. \onlinecite{Ivankov03} and for useful discussions. We are pleased 
to acknowledge conversations with M. Gruebele, V. Munoz and 
W. A. Eaton on the experimental results of the determination of folding
prefactors. Mai Suan Li would like to thank the hospitality of IPST,
where part of this work has been done.  This work was
supported in part by a KBN grant and the National Science Foundation grant
(NSF CHE-0209340).

\newpage

Table 1. Estimates of the folding and unfolding prefactors$^a$

\begin{tabular}{@{\extracolsep{15pt}}lrrrrrrr}\hline
protein$^b$ & $\beta \Delta G^c$ & $\tau_F^d$ & $\tau_U^e$ & $\tau_F^0$$^f$
            & $\tau_F^0$$^g$ & $\tau_U^0$$^h$ & $\tau_U^0$$^i$ \\ \hline
TI I27 (89)  & 12.7  & 0.0313  & 2041  & 0.974  & 23.9  & 0.194 & 4.76  \\
TWIg18$^\prime$(93) &  6.9  & 0.667   & 3571  & 16.5   & 412   & 89.0  & 2220 \\
CD2d1(98)           & 11.5  & 0.0556  & 588   & 1.04   & 26.4  & 0.111  & 2.83
\\
TNfn3 (92)          &  9.1  & 0.344   & 2174  & 9.00   & 224   & 6.35   & 158 \\
FNfn10 (96)         & 15.9  & 0.00417 & 4348  & 0.0870 & 2.20  & 0.0113 & 0.285
\\
CspB ($B.subtilis$)(67)    & 4.6 & 0.00145  & 0.101 & 0.178  & 3.82 & 0.125 & 2.68
   \\
CspB ($B.caldolyticus$)(66)& 8.1 & 0.000730 & 1.56  & 0.0960 & 2.04 & 0.0623& 1.32
 \\
CspB ($T. maritima$)(68)   & 10.6& 0.00177  & 55.6  & 0.203  & 4.40 & 0.159 & 3.44
  \\\hline

\end{tabular}

\noindent
(a) Data for the first five proteins are from \cite{Clarke99}
and the data for CspB proteins are from \cite{Perl98}\\
(b) Numbers in parenthesis are the values of $N$\\
(c) Free energy of stability extrapolated to zero denaturant concentration\\
(d) Folding times in seconds\\
(e) Unfolding times in seconds\\
(f) Folding prefactor (in units of $\mu s$) calculated using
$\tau ^0_F = \tau_F \exp(-1.1 N^{1/2})$\\
(g) Folding prefactor (in units of $\mu s$) calculated using
$\tau ^0_F = \tau_F \exp(-0.36 N^{2/3})$\\
(h) Unfolding prefactor (in $\mu s$) calculated using
$\tau_U^0 = \tau_U \exp(-1.1 N^{1/2} - \Delta G/(k_BT))$ \\
(i) Unfolding prefactor (in $\mu s$) calculated using
$\tau_U^0 = \tau_U \exp(-0.36 N^{2/3} -  \Delta G/(k_BT))$

\newpage

\begin{center}
{\bf Figure captions}
\end{center}

\vspace{0.3cm}

Fig. (1) (a) Temperature dependence of $d<\chi>/dT$ for the lattice
sequence with $N=64$. The
folding transition temperature is identified with the peak in
$d<\chi>/dT$. The full width at half-maximum is indicated by $\Delta
T$. (b) The dependence of $\Delta T/T_F$ as a function of $N$. The
straight line gives the fit $\Delta T/T_F \sim N^{\lambda}$ with $\lambda =
1.2 \pm 0.1 $.

Fig. (2) (a) Dependence of $ F/k_B T$ ($F$ is the free
energy of a sequence) as a function of the presumed reaction
coordinate $Q$, the number of native contacts, for one of the $N=64$ Go
sequences. The unfolding and refolding barriers are extracted from the
free energy profile as indicated. Panel (b)
shows the fit $\Delta F^\ddagger_F/k_B T_F \sim \text{ln} N$.
Panels (c) and (d)
correspond to the fits 
$\Delta F^\ddagger_F/k_B T_F \sim N^\beta$ with $\beta =0.5$ and
2/3, respectively. The results were computed for $N= 18(20), 27(17)$, 36(18),
48(18), 64(15), and 80(12), where the number in parenthesis refers to
the number of sequences used for averaging  $\Delta F^\ddagger_F/k_B
T_F$. Similar scaling with $N$ is obtained for $\Delta F^\ddagger_U/k_B
T_F$.

Fig. (3) Fits of ln $k_F$ as a function of $N$ for the dataset of 57
proteins and peptides taken from ref. \cite{Ivankov03}. Cross and
hexagon symbols 
correspond to three and two state folders, respectively. 
(a) The fit based on
ln $k_F \sim N^{\frac{1}{2}}$. The straight line is $y=-1.1x+14.7$ and the
correlation coefficient is 0.71. (b)  The fit based on ln $k_F \sim
N^{\frac{2}{3}}$. The straight line is  $y=-0.36x+11.7$ and the correlation
coefficient is 0.70. (c) Fits of ln $k_F \sim \text{ln} N$ gives
$y=-5.5x+28.5$ with the correlation coefficient of 0.72. 
(d) Variation of the correlation coefficient with
$\beta$. The correlation becomes weaker at  $\beta > 2/3$.

\newpage

\end{document}